\documentclass[a4paper,12pt]{article}
\usepackage[utf8x]{inputenc}

\usepackage{amsmath}
\usepackage{amsthm}
\usepackage{amssymb}
\usepackage{graphicx}
\usepackage{epsfig}
\usepackage{verbatim}
\usepackage{slashed}
\usepackage{color}

\newcommand{\eeVVV}{$e^-e^+\to VVV$~}
\newcommand{\eevvv}{$e^-e^+\to VVV$ with $V=\gamma,Z$~}
\newcommand{\eerrr}{$e^-e^+\to \gamma\gamma\gamma$~}
\newcommand{\eerrz}{$e^-e^+\to \gamma\gamma Z$~}
\newcommand{\eerzz}{$e^-e^+\to \gamma ZZ$~}
\newcommand{\eezzz}{$e^-e^+\to ZZZ$~}
\newcommand{\rrr}{$\gamma\gamma\gamma$~}
\newcommand{\rrz}{$\gamma\gamma Z$~}
\newcommand{\rzz}{$\gamma ZZ$~}
\newcommand{\zzz}{$ZZZ$~}

\title{Neutral Triple Gauge Boson production in the large extra dimensions model at linear colliders}
\author{
Sun Hao$^a$\footnote{haosun@mail.ustc.edu.cn}, Zhou Ya-Jin$^b$\footnote{zhouyj@sdu.edu.cn}\\
{\small $^{a}$ School of Physics and Technology, University of
Jinan, Jinan 250022, Shandong Province, P.R.China} \\
{\small $^{b}$ School of Physics, Shandong University, Jinan
250100, Shandong Province, P.R.China} }
\date{}

\begin{document}
\maketitle

\begin{abstract}

We consider the neutral triple gauge boson production process in the
context of large extra dimensions (LED) model including the
Kaluza-Klein (KK) excited gravitons at future linear colliders, say
ILC(CLIC). We consider $\gamma\gamma\gamma, \gamma\gamma Z, \gamma Z
Z$ and $ZZZ$ production processes, and analyse their impacts on both
the total cross section and some key distributions. These processes
are important for new physics searches at linear colliders. Our
results show that KK graviton exchange has the most significant
effect on \eerzz among the four processes with relatively small
$M_S$, while it has the largest effect on \eerrr with larger $M_S$.
By using the neutral triple gauge boson production we could set the
discovery limit on the fundamental Plank scale $M_S$ up to around
6-9 TeV for $\delta$ = 4 at the 3 TeV CLIC.
\end{abstract}


\baselineskip=0.32in
\newpage
\section{Introduction}
The hierarchy problem of the standard model (SM) strongly suggests
new physics at TeV scale, and the idea that there exists extra
dimensions (ED) which first proposed by Arkani-Hamed, Dimopoulos,
and Dvali\cite{ADD} might provide a solution to this problem. They
proposed a scenario in which the SM field is constrained to the
common 3+1 space-time dimensions (``brane"), while gravity is free
to propagate throughout a larger multidimensional space $D=\delta+4$
(``bulk"). The picture of a massless graviton propagating in D
dimensions is equal to the picture that numerous massive
Kaluza-Klein (KK) gravitons propagate in 4 dimensions. The
fundamental Planck scale $M_S$ is related to the Plank mass scale
$M_{Pl}=G_N^{-1/2}=1.22\times10^{19}~{\rm GeV}$ according to the
formula $M^2_{Pl}=8\pi M^{\delta+2}_{S} R^\delta$ , where $R$ and
$\delta$ are the size and number of the extra dimensions,
respectively. If $R$ is large enough to make $M_S$ on the order of
the electroweak symmetry breaking scale ($\sim 1~ {\rm TeV}$), the
hierarchy problem will be naturally solved, so this extra dimension
model is called the large extra dimension model (LED) or the ADD
model. Postulating $M_S$ to be 1 TeV, we get $R\sim 10^{13}~{\rm
cm}$ for $\delta=1$, which is obviously ruled out since it would
modify Newton's law of gravity at solar-system distances; and we get
$R\sim 1~{\rm mm}$ for $\delta=2$, which is also ruled out by the
torsion-balance experiments\cite{Kapner:2006si}. When $\delta \geq
3$, where $R < 1~{\rm nm}$, it is possible to detect graviton signal
at high energy colliders.

At colliders, exchange of virtual KK graviton or emission of a real
KK mode could give rise to interesting phenomenological signals at
TeV scale\cite{ADD:Gian,ADD:HanTao}. Virtual effects of KK modes
could lead to the enhancement of the cross section of pair
productions in processes, for example, di-lepton, di-gauge boson
($\gamma\gamma$, $ZZ$, $W^+W^-$), dijet, $t\bar{t}$ pair, HH
pair\cite{ADDvirtualA, ADDvirtualB, ADDvirtualC, ADDvirtualD,
ADDvirtualE, ADDvirtualF, ADDvirtualG} etc. The real emission of a
KK mode could lead to large missing $E_T$ signals viz. mono jet,
mono gauge boson\cite{ADD:Gian, ADD:HanTao, ADDrealA, ADDrealB} etc.
The CMS Collaboration has performed a lot of search for LED on
different final states at $\sqrt{s}=7$
TeV\cite{CMS:LED1,CMS:LED2,CMS:LED3}, and they set the most
stringent lower limits to date to be $2.5~{\rm TeV} < M_S <3.8~{\rm
TeV}$ by combining the diphoton, dimuon and dielectron channels.

Studies for LED have been extended to three body final state
processes in recent years. Triple gauge bosons productions in the SM
are important because they involve 3-point and 4-point gauge
couplings in the contributing diagrams, which allow for restrictive
tests of triple and quartic vector boson coupling. And also they
might contribute backgrounds to new physics beyond the SM.
Furthermore they are sensitive to new physics. This kind of
processes have been studied at LO\cite{VVV:Golden, VVV:Barger,
VVV:HanT} and NLO\cite{eevvv_NLO_Fawzi, eevvv_NLO_Su, eevvv_NLO_Sun,
eevvv_NLO_Binoth, eevvv_NLO_Lazopoulos, eevvv_NLO_Bozzi} in the SM,
and virtual graviton exchange effects to these processes within the
LED model at LHC are studied recently\cite{VVV:ED:Kumar}. Linear
colliders have more advantage in testing extra dimensions than LHC
for the following reasons. First, even though the LHC has much
higher center-of-mass (c.m.s.) energy than linear colliders, the
theoretical amplitude at LHC is hampered by the unitary constraint
$\sqrt{\hat{s}}<{\rm M_S}$. Second, linear colliders have cleaner
environment than LHC, so it's much easier to select the ED signals.
The capabilities of the planned International Linear Collider (ILC)
and Compact Linear Collider (CLIC) for precision Higgs studies are
well documented\cite{ILCforHiggs, CLICforHiggs}. They will also
provide opportunities for the search for new physics beyond SM. So
in this paper we consider the triple gauge bosons production at ILC
and CLIC \eeVVV within the LED model, where we restrict V to be
neutral gauge boson ($V=\gamma,Z$). The following four final states
are the subject of this analysis: (i) $\gamma\gamma\gamma$ (ii)
$\gamma\gamma Z$ (iii) $\gamma Z Z$ (iv) $ZZZ$. The case where $V =
W^\pm$ is in prepare and will be part of a different paper.

This paper is organized as follows: in section 2 we present the
analytical calculation of the processes mentioned above with a brief
introduction to the LED model, section 3 is arranged to present the
numerical results of our studies, and finally we summarize the
results in the last section.

\section{Theoretical Framework}

\par
In this section we give the analytical calculations of the process
\eeVVV with $V =\gamma,Z$ at linear colliders in the LED model. In
our calculation we use the de Donder gauge. The relevant Feynman
rules involving graviton in the LED model can be found in
Ref.\cite{ADD:HanTao}. We denote the process as:
\begin{eqnarray}
e^-(p_1) + e^+(p_2) \rightarrow V(p_3) + V(p_4) + V(p_5)
\end{eqnarray}
where $p_1$, $p_2$ and $p_3$, $p_4$, $p_5$ represent the momenta of
the incoming and outgoing particles respectively.

\begin{figure}[hbtp]
\includegraphics[angle=0,scale=1]{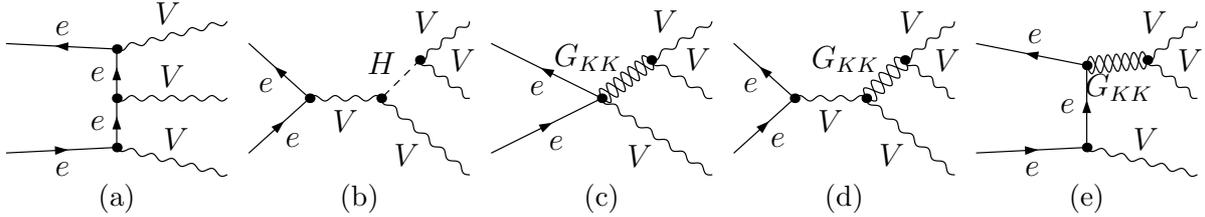}
\caption{\label{feynfig} Feynman diagrams for \eeVVV process with
$V=\gamma,Z$. (a,b) are the diagrams in the SM, and (c,d,e) are the
diagrams in the LED model.}
\end{figure}

In Fig.\ref{feynfig} we display the Feynman diagrams for this
process in both the SM and LED model, among which (a) and (b) are SM
diagrams and (c)$\sim$(e) are LED diagrams. We have neglected the
Higgs coupling to electrons because the Yukawa coupling is
proportional to the fermion mass. But the Higgs coupling to Z
bosons, which appears in \eezzz process (Fig.\ref{feynfig}(b)),
can't be neglected because of its large contribution, e.g., with
$\sqrt{s}$ to be $300-3000~{\rm GeV}$, the cross section for
Fig.\ref{feynfig}(b) is about $26\% - 9\%$ of the total SM cross
section. After considering all possible permutations, we have 12 SM
diagrams for the $ZZZ$ production process and 9 SM diagrams for the
other three processes, and we have 12 LED diagrams for the \rrr and
$ZZZ$ processes and 4 LED diagrams for \rrz and \rzz processes.

\par
In our calculation we consider both the spin-0 and spin-2 KK mode
exchange effect. The spin-0 states only couple through the dilaton
mode, which have none contribution to \rrr and \rrz processes and
could contribute to \rzz and \zzz production processes through
couplings to massive gauge bosons. However the cross sections coming
from the dilaton mode are so small that can be neglected, e.g., they
are at most about $10^{-12}$ and $10^{-8}$ times of the total cross
sections for \rzz and \zzz production processes, respectively. So we
focus our study on the spin-2 component of the KK states.

\par
The couplings between gravitons and SM particles are proportional to
a constant named gravitational coupling $\kappa \equiv \sqrt{16 \pi
G_N}$, which can be expressed in terms of the fundamental Plank
scale $M_S$ and the size of the compactified space R by
\begin{eqnarray}
  \kappa^2 R^{\delta} = 8 \pi (4 \pi)^{\delta/2} \Gamma(\delta/2) M_S^{-(\delta+2)}
\end{eqnarray}

In practical experiments, the contributions of the different
Kaluza-Klein modes have to be summed up, so the propagator is
proportional to $i/(s_{ij}-m^2_{\vec{n}})$, where
$s_{ij}=(p_i+p_j)^2$ and $m_{\vec{n}}$ is the mass of the KK state
$\vec{n}$. Thus, when the effects of all the KK states are taken
together, the amplitude is proportional to $\sum\limits_{\vec{n}}
\frac{i}{s_{ij}-m^2_{\vec{n}}+i\epsilon}=D(s)$. If $\delta \geq 2$
this summation is formally divergent as $m_{\vec{n}}$ becomes large.
We assume that the distribution has a ultraviolet cutoff at
$m_{\vec{n}}\sim M_S$, where the underlying theory becomes manifest.
Then $D(s)$ can be expressed as:
\begin{eqnarray}
 D(s) =\frac{1}{\kappa^2}
\frac{8\pi}{M_S^4}(\frac{\sqrt{s}}{M_S})^{\delta-2}[\pi + 2i
I(M_S/\sqrt{s})].
\end{eqnarray}
The imaginary part I($\Lambda/\sqrt{s}$) is from the summation over
the many non-resonant KK states and  its expression can be found in
Ref.\cite{ADD:HanTao}. Finally the KK graviton propagator after
summing over the KK states is:
\begin{eqnarray}
\label{prop}
\tilde{G}^{\mu\nu\alpha\beta}_{KK}=D(s)
 \left(\eta_{\mu\alpha}\eta_{\nu\beta} + \eta_{\mu\beta}\eta_{\nu \alpha}
- \frac{2}{D-2}\eta_{\mu\nu}\eta_{\alpha\beta}\right)
\end{eqnarray}

Using the Feynman rules in the LED model and the propagator given by
Eq.(\ref{prop}), we can get the amplitudes for the  virtual KK
graviton exchange diagrams in Fig.\ref{feynfig}. The total amplitude
can be obtained by adding these LED amplitudes together with the SM
ones. The total cross section can be expressed as the integration
over the phase space of three-body final state:
\begin{eqnarray}
\sigma_{tot}&=&\frac{(2\pi)^4 }{4|\vec p_1|\sqrt{s}}\int d\Phi_3
\overline{\sum}|{\cal M}_{tot}|^2.
\end{eqnarray}
where $\overline{\sum}$ represents the summation over the spins of
final particles and the average over the spins of initial particles.
The phase-space element $d\Phi_3$ is defined by
\begin{eqnarray}
{d\Phi_3}=\delta^{(4)} \left( p_1+p_2-\sum_{i=3}^5 p_i \right)
\prod_{j=3}^5 \frac{d^3 \textbf{\textsl{p}}_j}{(2 \pi)^3 2 E_j}.
\end{eqnarray}

\section{Numerical Results}
\subsection{Input parameters and kinematical cuts}
We use FeynArts and FormCalc package\cite{FeynArts,FormCalc} to
generate and reduce the amplitudes and then implement numerical
calculation. We use BASES\cite{BASES} to perform the phase space
integration and CERN library to display the distributions. The SM
parameters are taken as follows\cite{ParticleDataGroup}:
\begin{eqnarray}\nonumber
&&m_Z=91.1876~{\rm GeV},~ m_W=80.399~{\rm GeV}, ~m_e=0.511~{\rm
MeV},
\\ \nonumber
 &&\alpha(m^2_Z)=1/127.934, ~m_{H}=125~{\rm GeV}\cite{SMHiggs125GeV_ATLAS, SMHiggs125GeV_CMS}.
\end{eqnarray}

We take the cuts on final particles as:
\begin{eqnarray}
 p_T^{\gamma,Z} \geqslant 15~{\rm GeV}, \ \ \eta^{\gamma,Z}\leqslant 2.5,\ \ R_{\gamma\gamma}\geqslant0.4
\end{eqnarray}
where R is defined as $R=\sqrt{(\Delta\phi)^2 + (\Delta\eta)^2}$,
with $\Delta\phi$ and $\Delta\eta$ denoting the separation between
the two particles in azimuthal angle and pseudo-rapidity
respectively.

\subsection{Total Cross sections}

\begin{figure}[h]
\centering
\scalebox{0.9}[0.9]{\includegraphics*[68pt,42pt][594pt,418pt]{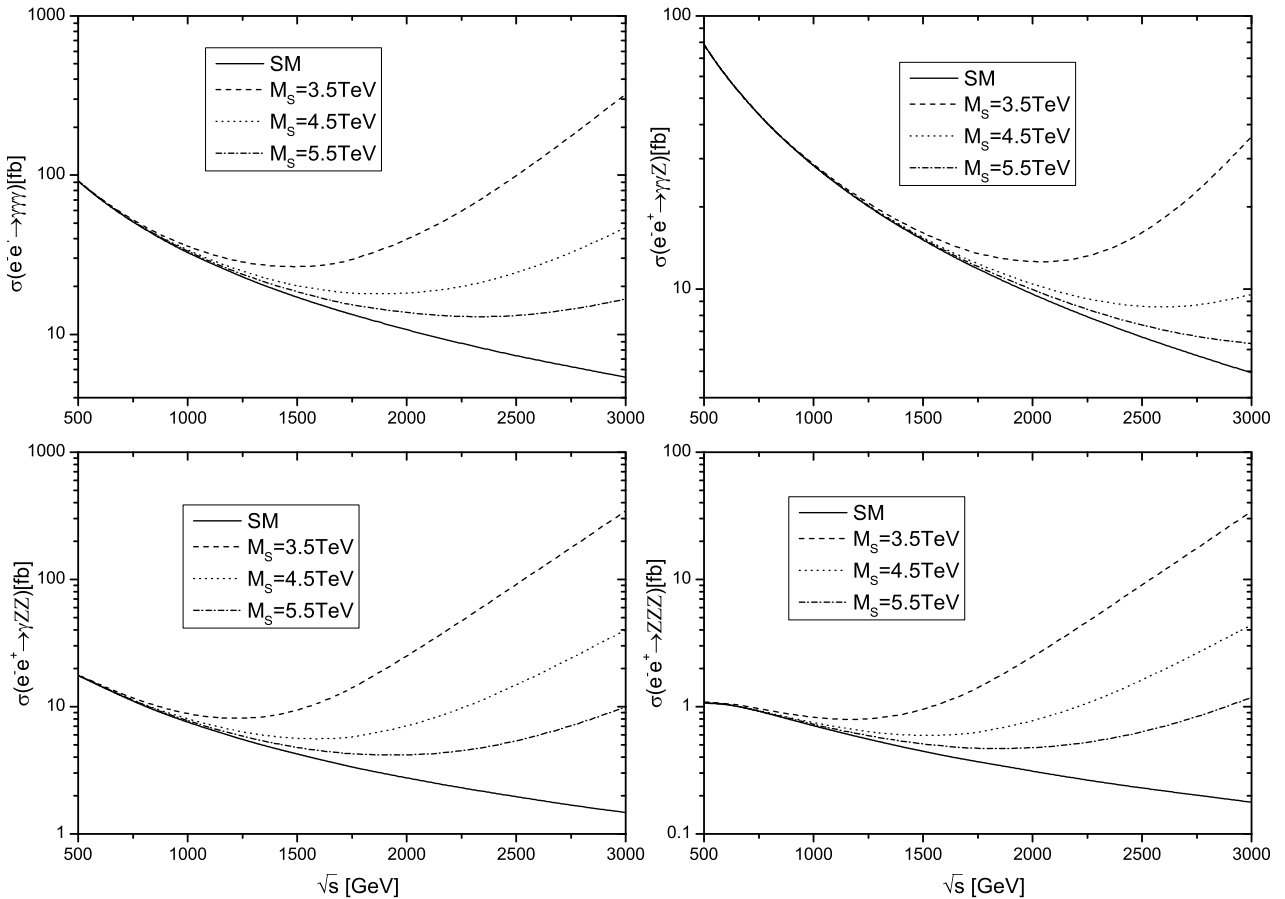}}
\caption{\label{runS} The cross sections for the process \eevvv in
the SM and LED model as the function of $\sqrt{s}$ with $M_{s}=$
3.5, 4.5, 5.5 TeV and $\delta=3$.}
\end{figure}

In Fig.\ref{runS} we present the cross sections for
$e^{-}e^{+}\rightarrow \gamma\gamma\gamma,\gamma\gamma Z,\gamma ZZ\
{\rm and}\ ZZZ$ processes as the functions of $\sqrt{s}$ for
$\delta=3$ with different values of $M_S$. The solid lines are the
SM results. The dashed, dotted and dot-dashed lines are
corresponding to the cross sections in the LED model for $M_S=3.5$
TeV, 4.5 TeV and 5.5 TeV respectively. When $\sqrt{s}$ is less than
about 1 TeV, the curves for the total cross sections including the
LED effect seem to be overlapped with that in the SM, then the LED
effect becomes significant with the increment of $\sqrt{s}$. When
$M_S=3.5$ TeV, the \eerzz process has the most significant LED
effect among the four processes considered in this paper, while the
\eerrz process has the least. When $\sqrt{s}=500~{\rm GeV}$, the
cross sections for \rrr and \rrz production processes are
comparable, which are 91 fb and 78 fb respectively, because they
have comparable phase-space. While the cross sections for \rzz and
\zzz processes are much smaller due to the less phase-space, which
are 18 fb and 1 fb, respectively. With $\sqrt{s}$ increase to be 3
TeV, the cross section for \rzz are enhanced to 347 fb, which is
even larger than the \rrr process (323 fb), and the cross sections
for \zzz process is enhanced to 34 fb, which is comparable to the
\rrz process (36 fb). With larger $M_S$ value (5.5 TeV), the LED
contribution to \rrr production process will exceed \rzz, that's why
\eerrr process puts the highest limits on $M_S$, as we will see
later.

\par
\begin{figure}[htbp]
\centering
\scalebox{0.9}[0.9]{\includegraphics*[68pt,38pt][594pt,418pt]{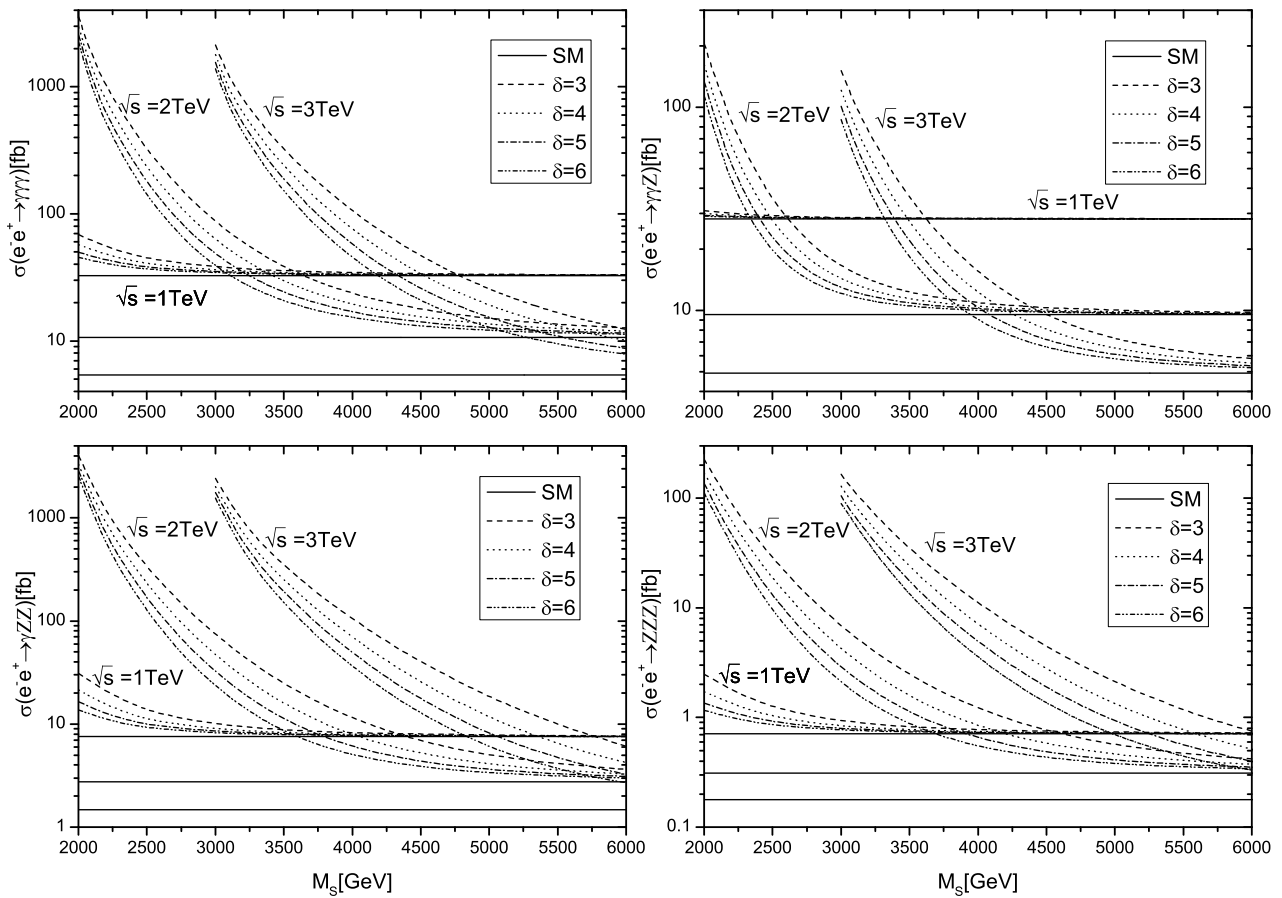}}
\caption{\label{runMs} The cross sections for the process
$e^{-}e^{+}\rightarrow VVV$ in the SM and LED model as the function
of $M_S$ with $\sqrt{s}=$ 1, 2, 3 TeV and $\delta=3,~4,~5,~6$.}
\end{figure}

In Fig.\ref{runMs} we present the dependence of the cross section on
energy scale $M_S$ with $\sqrt{s}=1$ TeV, 2 TeV and 3 TeV
respectively. In each figure of Fig.\ref{runMs}, we present the
curves for the cross sections with the extra dimension $\delta$
value being 3, 4, 5 and 6 separately. The solid straight lines,
which are independent of $M_S$, are the SM results, and the dashed,
dotted, dash-dotted and dash-dot-dotted lines are the cross sections
for $\delta$=3, 4, 5 and 6 respectively. It's clear that for a given
value of $\delta$, the cross section decreases rapidly with the
increment of $M_S$, and finally approaches to its corresponding SM
result. We can see again that the virtual KK graviton exchange
contribution decreases with the increment of the $\delta$ value. The
LED effect on the cross sections with $\sqrt{s}=1$ TeV is too small
to be detected, especially for \eerrz process, which is coincidence
with Fig.\ref{runS}. If we got high enough c.m.s energy, say 3 TeV,
the cross sections would be very significant when $M_S$ is not very
large. Even with $M_S=6$ TeV, the cross sections are still several
times of the SM ones.
\par

\subsection{Distributions}
\par
The distributions of the Gauge boson pair invariance mass $M_{VV}$
($VV=\gamma\gamma,~ZZ$) and the Gauge boson transverse momentum
$p_{T}^{V}$ as well as their rapidity $y^{V}$ at the 3 TeV CILC, are
shown in Fig.4-6. The results are for $M_S=6.5$ TeV at the fixed
value 4 for the number of extra dimensions and obtained by taking
the input parameters mentioned above.

\subsection*{$e^{-}e^{+}\rightarrow \gamma\gamma\gamma$}
Before selecting our event samples for triple $\gamma$ production,
we order the photons on the basis of their transverse momentum i.e.,
$p_T^{\gamma_1}\geq p_T^{\gamma_2} \geq p_T^{\gamma_3}$. For
$e^{-}e^{+}\rightarrow \gamma\gamma\gamma$, we are interested in the
$p_{T}^{\gamma}$ and $y^{\gamma}$ distribution which are displayed
in the left and right panel in Fig.4, respectively. The solid,
dashed and dotted lines refer to $\gamma_1$, $\gamma_2$ and
$\gamma_3$, respectively. In high $p_{T}^{\gamma_1}$ and
$p_{T}^{\gamma_2}$ region, the LED effect dominant the total
(SM+LED) distribution, because more KK modes contribute with the
increase of $p_T$. Difference can be found for the
$p_{T}^{\gamma_3}$ production, although it's still enhanced by the
LED effects, it's low $p_T$ region is dominant while in high $p_T$
region it becomes much smaller. Rapidity distribution of the related
photon has been shown in the right panel in Fig.4. As we can see,
the rapidity distributions in the LED model show significantly peaks
around $y = 0$, which implies the large contributions at high
$p_{T}^{\gamma_{1,2}}$ region. Compare with the $\gamma_1$ and
$\gamma_2$ distribution, the $y$ distribution of $\gamma_3$ seems
much flatter.

\begin{figure}[hbtp]
\centering
\scalebox{0.44}[0.44]{\includegraphics*[7pt,17pt][512pt,416pt]{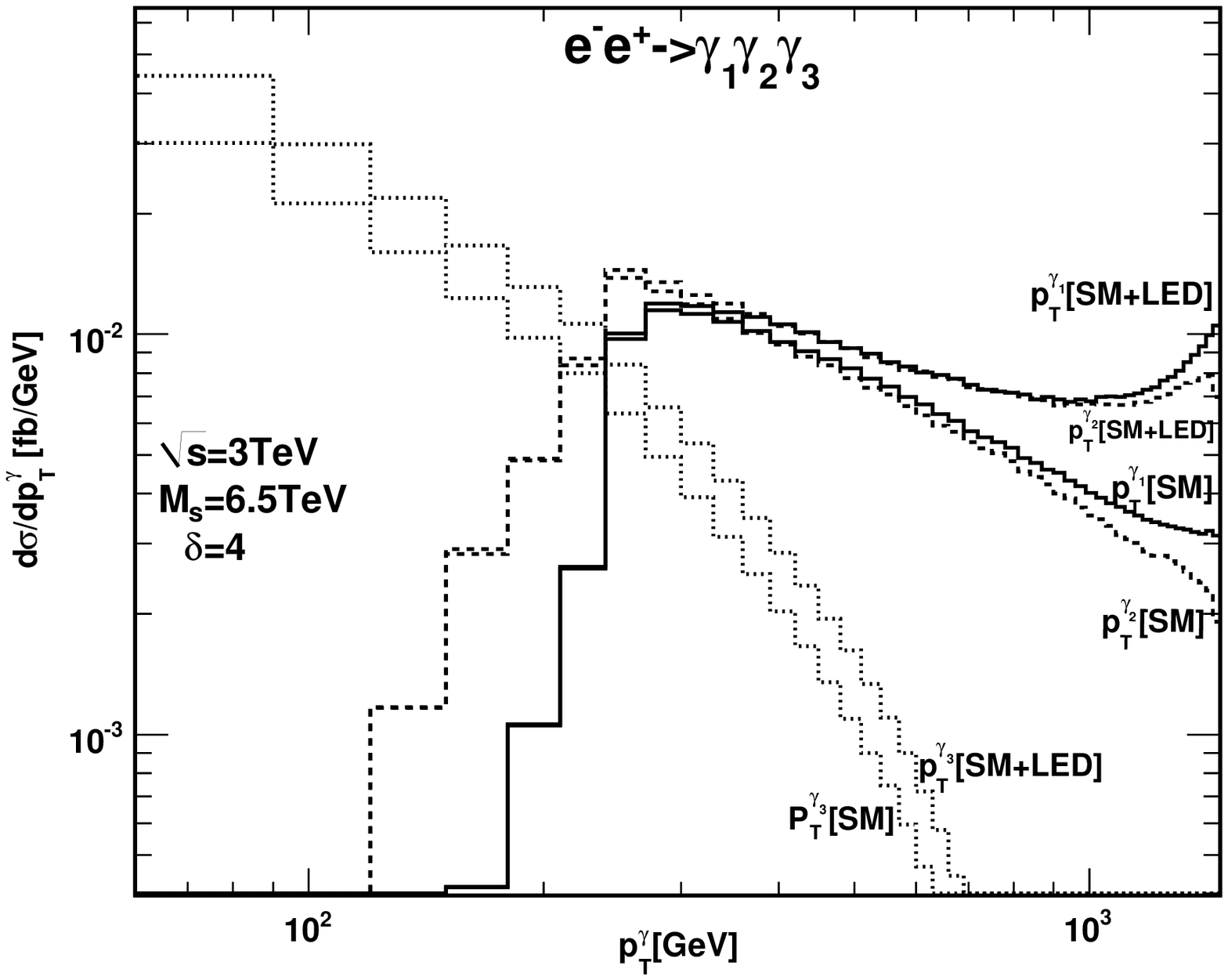}}
\scalebox{0.44}[0.44]{\includegraphics*[7pt,17pt][512pt,416pt]{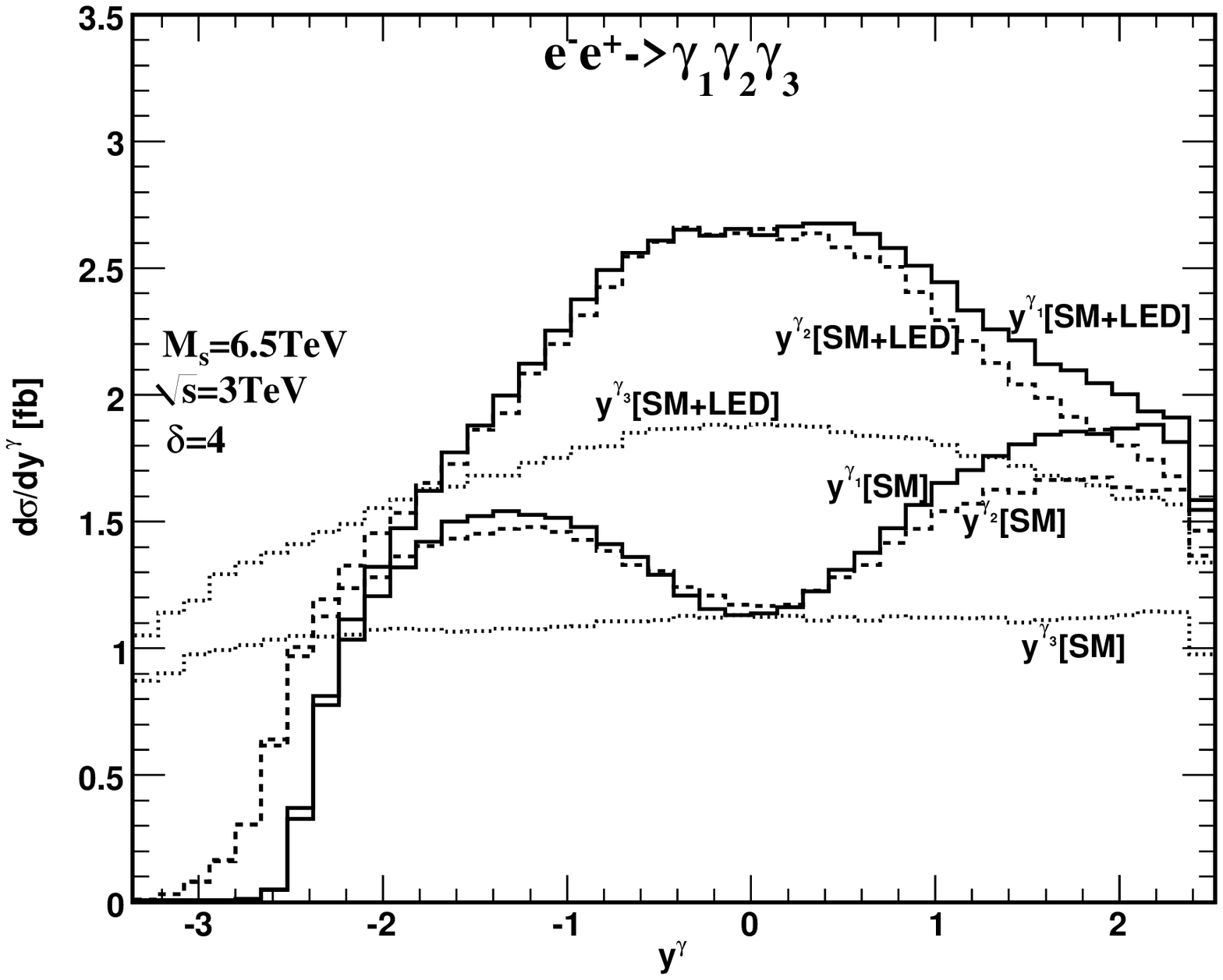}}
\caption{\label{fig4} The transverse momentum($p_T$) and
Rapidity($y$) distribution of photons for the process
$e^{-}e^{+}\rightarrow \gamma\gamma\gamma$, on the basis of their
transverse momentum $p_T^{\gamma_1}\geq p_T^{\gamma_2} \geq
p_T^{\gamma_3}$ with $M_S=$ 6.5TeV, $\sqrt{s}=$3 TeV and $\delta=4$.
The solid, dashed and dotted lines refer to
$p_T^{\gamma_1}$($y^{\gamma_1}$), $p_T^{\gamma_2}$($y^{\gamma_2}$)
and $p_T^{\gamma_3}$($y^{\gamma_3}$).}
\end{figure}

\subsection*{$e^{-}e^{+}\rightarrow ZZZ$}
Similar to the $e^{-}e^{+}\rightarrow \gamma\gamma\gamma$
production, triple Z bosons final particles are classified in such a
way that $p_T^{Z_1}\geq p_T^{Z_2} \geq p_T^{Z_3}$. Similar
conclusion can be found for the $e^{-}e^{+}\rightarrow ZZZ$
production. It's not strange that the signal of triple $\gamma$
signal is larger than the triple Z production since the three Z
bosons suppress the phase space integration extremely, so that the
total cross sections as well as the distributions become smaller as
can be seen in Fig.4 and Fig.5, the peak is around 0.0006 fb/GeV for
$p_{T}^{Z_1}$ compared to 0.01 fb/GeV for $p_{T}^{\gamma_1}$ in the
high $p_T$ region. For the $y$ distributions, the peaks for the
$ZZZ$ production are narrower than the $\gamma\gamma\gamma$
distributions, however, the conclusion is the same that the rapidity
distributions in the LED model show significant peaks around $y =
0$.

\begin{figure}[hbtp]
\centering
\scalebox{0.44}[0.44]{\includegraphics*[7pt,17pt][512pt,416pt]{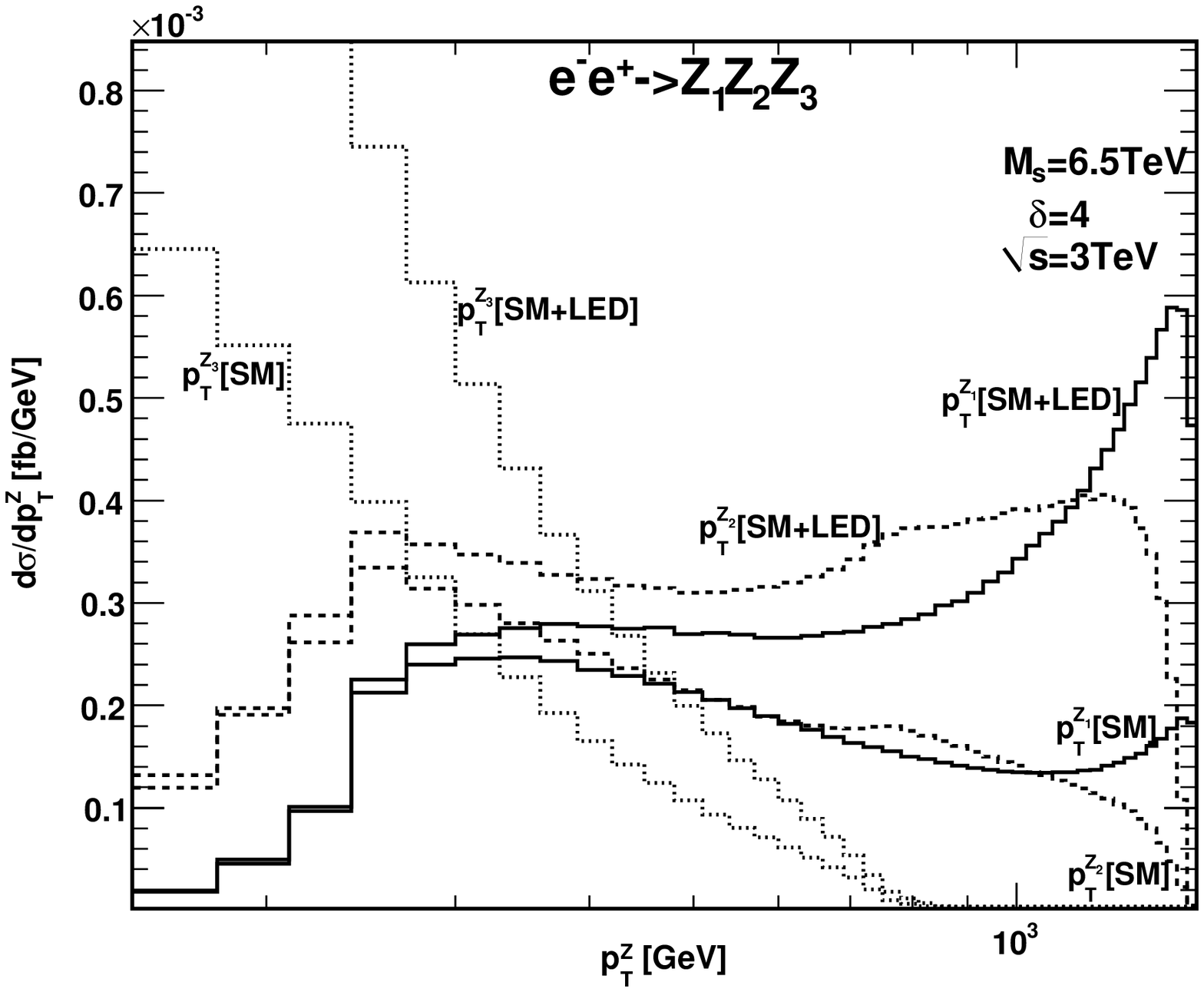}}
\scalebox{0.44}[0.44]{\includegraphics*[7pt,17pt][512pt,416pt]{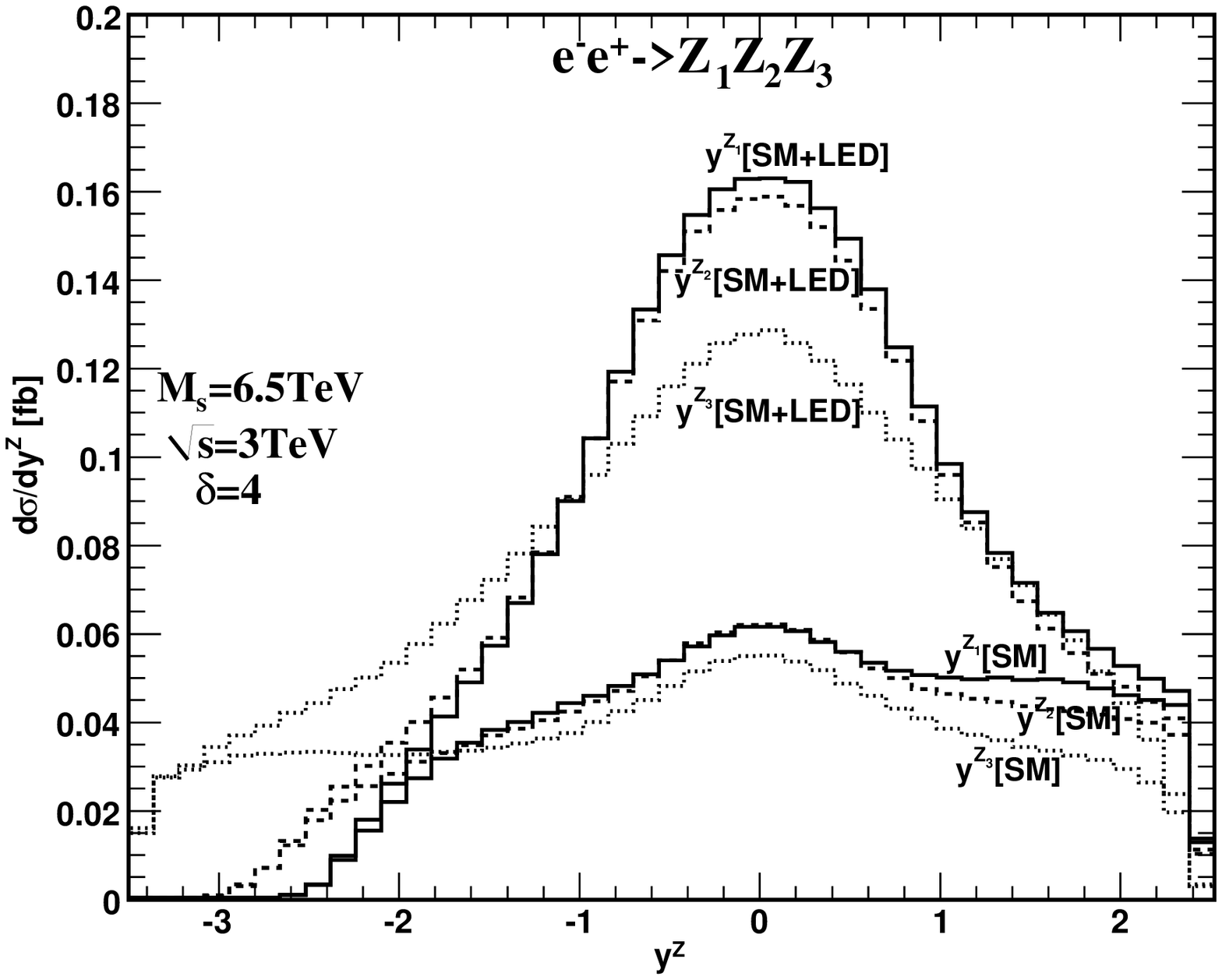}}
\caption{\label{fig5} The transverse momentum($p_T$)and
Rapidity($y$) distribution of $Z$ bosons for the process
$e^{-}e^{+}\rightarrow ZZZ$, on the basis of their transverse
momentum $p_T^{Z_1}\geq p_T^{Z_2} \geq p_T^{Z_3}$ with $M_S=6.5$
TeV, $\sqrt{s}=3$ TeV and $\delta=4$. The solid, dashed and dotted
lines refer to $p_T^{Z_1}$($y^{Z_1}$), $p_T^{Z_2}$($y^{Z_2}$) and
$p_T^{Z_3}$($y^{Z_3}$).}
\end{figure}

\subsection*{$e^{-}e^{+}\rightarrow \gamma\gamma Z$ and $e^{-}e^{+}\rightarrow \gamma Z Z$}
Now let's see the distributions for the $e^{-}e^{+}\rightarrow
\gamma\gamma Z$ and $e^{-}e^{+}\rightarrow \gamma Z Z$ productions.
The photon pair decay of the KK graviton is one of the clean decay
modes, so the distribution of the invariant mass of the photon pair
($M_{\gamma\gamma}$) is a useful observable for \eerrz. An obvious
enhancement on the tail of this distribution makes such region of
extreme interest. Typically, we find that the KK modes dominate over
SM contribution for larger values of invariant masses (say above 1
TeV for a given set of $M_S$ and $\delta$ values, here we give
$M_S=4.5$ TeV and $\delta=4$) of photon pairs indicating the
observable nature of the signal, see the two dotted line in Fig.6.
The upper and lower ones refer to the SM predict and SM+LED effects.
For the process $e^{-}e^{+}\rightarrow \gamma Z Z$, it is similar to
$\gamma\gamma Z$ production process, and in this case, the invariant
mass of Z boson pair($M_{ZZ}$) is a useful observable. We thus
display it in Fig.6, see the solid lines. These two solid line
branch at about the invariance mass 1 TeV, and the upper and lower
ones present the SM and SM+LED effects, respectively.

\begin{figure}[hbtp]
\centering
\includegraphics[angle=0,scale=0.5]{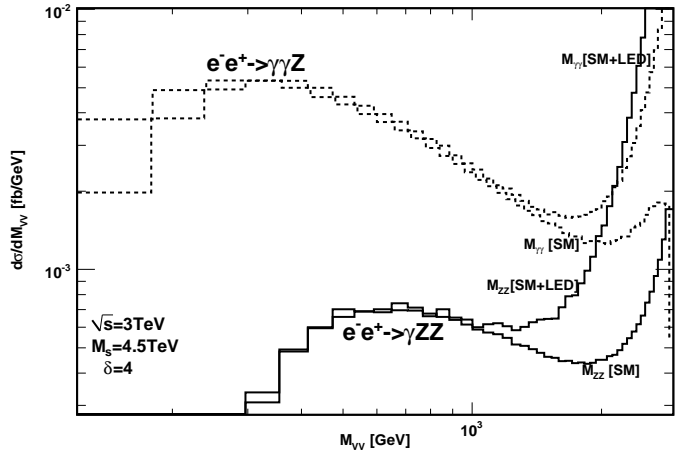}
\caption{\label{fig6} Invariant mass distribution of
$M_{\gamma\gamma(ZZ)}$ for $e^{-}e^{+}\rightarrow \gamma\gamma
Z$($e^{-}e^{+}\rightarrow \gamma ZZ$) for $M_S=4.5$ TeV,
$\sqrt{s}=3$ TeV and $\delta=4$.}
\end{figure}

\par
It is clear that if the deviation of the cross section from the SM
prediction is large enough, the LED effects can be found. We assume
that the LED effects can and cannot be observed, only
if\cite{sunhaorrttH}
\begin{eqnarray}
\Delta\sigma=|\sigma_{tot}-\sigma_{SM}|\geq  \frac{5\sqrt{ {\cal
L}\sigma_{tot}}}{{\cal L}} \equiv 5\sigma
\end{eqnarray}
and
\begin{eqnarray}
\Delta\sigma=|\sigma_{tot}-\sigma_{SM}|\leq  \frac{3\sqrt{ {\cal
L}\sigma_{tot}}}{{\cal L}} \equiv 3\sigma
\end{eqnarray}

\begin{table}
\begin{center}
\begin{tabular}{l c c c c c r r r r r r }
\hline\hline
  $\sqrt{s}$    &&\multicolumn{2}{c}{1 TeV}&& \multicolumn{2}{c}{2 TeV}&& \multicolumn{2}{c}{3 TeV} \\ [0.5ex]
      && $5\sigma$ & $3\sigma$&& $5\sigma$ & $3\sigma$ && $5\sigma$ & $3\sigma$ &&  \\
\hline
$e^{-}e^{+}\rightarrow \gamma\gamma\gamma$ && 3599 & 4093  && 6371 & 7187 && 8906  & 10023  \\
$e^{-}e^{+}\rightarrow ZZZ$                && 2530 & 2858  && 4711 & 5312 && 6730 & 7584  \\
$e^{-}e^{+}\rightarrow \gamma\gamma Z$ &&  2052 & 2289  && 4007 & 4458 && 5823 & 6500   \\
$e^{-}e^{+}\rightarrow \gamma ZZ$ && 3406 & 3835  &&  6027 & 6757  && 8412  & 9421  \\
\hline\hline
\end{tabular}
\end{center}
\vspace*{-0.8cm}
\begin{center}
\begin{minipage}{14cm}
\caption{\label{tab3} The discovery ($\Delta\sigma \geq 5\sigma$)
and exclusion ($\Delta\sigma \leq 3\sigma$) LED model fundamental
scale ($M_S$) values for the \eeVVV processes at the $\sqrt{s}$ = 1,
2, 3 TeV ILC(CLIC). ${\cal L}= 300~ {\rm fb}^{-1}$, $\delta = 4$.}
\end{minipage}
\end{center}
\end{table}

Our final results show that by using the \rrr production we can set
the discovery limit on the fundamental Plank scale $M_S$ up to
3.1-9.9 TeV, depending on the extra dimension $\delta\subset[3,6]$,
with the luminosity 300 fb$^{-1}$ and the colliding energy 1-3 TeV.
For the other three final states \rrz, \rzz and \zzz, the limits are
1.8-6.3 TeV, 2.9-9.3 TeV and 2.2-7.4 TeV, respectively. To do a more
detailed description, in Table 1, we present the $5\sigma$ discovery
and $3\sigma$ exclusion fundamental scale $M_S$ values at the
ILC/CLIC with the luminosity 300 fb$^{-1}$ for $\delta=4$ . It shows
that compared to the other three channels, $\gamma\gamma\gamma$ can
set the discovery limit bounds much higher, up to 8.9 TeV. The
phenomenology of the neutral triple gauge boson production at the
near future is much richer at linear colliders, though its
production cannot give compete limits as, for example, dilepton
production gives, it's still very interesting and important.

\vspace*{0.4cm}
\begin{table}[h]
\begin{center}
\begin{tabular}{l c c c c c r r r r r r }
\hline\hline
&& ${\cal L}$ (${\rm fb}^{-1}$)\\
        && $5\sigma$ & $3\sigma$  \\
\hline
$pp\rightarrow \gamma\gamma\gamma$ && 960 & 1500    \\
$pp\rightarrow ZZZ$                && 21 & 21    \\
$pp\rightarrow \gamma\gamma Z$ &&  20 & 17     \\
$pp\rightarrow \gamma ZZ$ && 170 & 140    \\
\hline\hline
\end{tabular}
\end{center}
\vspace*{-0.8cm}
\begin{center}
\begin{minipage}{14cm}
\caption{\label{tab3} Integrated luminosity needed at 14 TeV LHC to
accomplish the discovery and exclusion bounds at a 1 TeV linear
collider, which are listed in the first two columns in Table 1,
using the neutral triple gauge boson production processes, with
$\delta=4$. }
\end{minipage}
\end{center}
\end{table}

To make a comparison with Ref.\cite{VVV:ED:Kumar}, we repeat the $pp
\to VVV$ ($V=\gamma, Z$) process at LHC, using the same parameters
and cuts with Ref.\cite{VVV:ED:Kumar}, and find that our results are
in good agreement with theirs. In Table 2 we list the integrated
luminosity the 14 TeV LHC needed to accomplish the discovery and
exclusion bounds at a 1 TeV LC (the first two column data listed in
Table 1), by using the corresponding $VVV$ production channels with
extra dimensions $\delta=4$. The table shows that with years of
collection of data, LHC could accomplish the discovery and exclusion
limits set by a 1 TeV LC, even for the most challenging channel
\rrr. While the limits set by a 2 or 3 TeV LC are much higher, and
the required amounts of data for matching these bounds are too large
to be a reasonable projection for the LHC reach.

\section{Summary and Conclusions}
\par
In a short summary, we calculate the neutral gauge boson production
processes $\gamma\gamma\gamma$, $\gamma\gamma Z$, $\gamma Z Z$ and
$ZZZ$ in the SM and LED model at ILC and CLIC. We investigate the
integrated cross sections, the distributions of some kinematic
variables $M_{VV}$, $p_T^V$ and $y^{V}$. The 5$\sigma$ discovery and
3$\sigma$ exclusion ranges for the LED parameters $M_S$ are obtained
and compared between different channels. It turns out that the
effects of the virtual KK graviton enhance the total cross sections
and differential distributions of kinematical observables generally.
Among the four processes we considered, \eerzz or \eerrr process has
the most significant LED effect with relatively small or large
$M_S$, respectively. While \eerrz has the least contribution from
LED diagrams. With the development of linear colliders, more
information related to LED effects can be obtained experimentally
through such important productions. At the 3 TeV CLIC, it is
expected that \rzz production can be used to explore a range of
$M_S$ values up to 7.3-9.3 TeV depending on the number of extra
dimensions. Through $\gamma\gamma\gamma$ production, we can extend
this search up to 9.9 TeV, While using $\gamma\gamma Z$ and $ZZZ$
productions, lower bounds on $M_S$ can be found, which are 6.3 TeV
and 7.4 TeV, respectively.

\section{Acknowledgments}
We would like to thank Prof. Zhang Ren-You for useful discussions.
Project supported by the National Natural Science Foundation of
China (Grant No.11147151, No.11205070, No.11105083, No.10947139 and
No.11035003), and by Shandong Province Natural Science Foundation
(No.ZR2012AQ017).

\end{document}